\documentclass[11pt,a4paper,eps]{article}

\def\be{\begin{equation}}
\def\ee{\end{equation}}
\def\bea{\begin{eqnarray}}
\def\eea{\end{eqnarray}}

\usepackage{fancyhdr}
\usepackage{amsfonts}
\usepackage{amsmath}
\usepackage{amssymb}
\usepackage{color}

\def\>#1{{\bf #1}}

\def\>#1{{\bf #1}}

\newcommand{\nn}{\nonumber}

\makeatletter
\@addtoreset{equation}{section}
\makeatother

\def\appendix#1
\renewcommand{\thesection}{\Alph{section}}

\def\<#1,#2>{\left\langle#1,#2\right\rangle}

\def\nn{\nonumber}

\def\be{\begin{equation}}
\def\ee{\end{equation}}
\def\bea{\begin{eqnarray}}
\def\eea{\end{eqnarray}}

\newcommand{\al}{\alpha}

\newcommand{\nab}[1]{\vec{x}\,\vec{\nabla}}

\usepackage{graphicx,amssymb}
\usepackage{amsmath}
\usepackage{graphicx}
\usepackage{amsfonts}
\usepackage{amssymb}
\begin{document}

\pagestyle{empty} \addtolength{\topmargin}{20mm}

\begin{center}
{\LARGE {\bf Generalizing the Noether theorem\\ for Hopf-algebra
spacetime symmetries} }
\end{center}

\vskip1.0 cm

\begin{center}
\textbf{Alessandra AGOSTINI$^a$, Giovanni AMELINO-CAMELIA$^{a,b}$, Michele ARZANO$^{c}$,\\
 Antonino MARCIAN\`O$^{a,b}$, Ruggero Altair~TACCHI$^{a}$} \\[0pt]
\vskip0.3 cm

\textit{$^a$Dipartimento di Fisica, Universit\`{a} di Roma ``La Sapienza'',P.le A. Moro 2, 00185 Roma, Italy}

\textit{$^b$INFN Sez.~Roma1, P.le A. Moro 2, 00185 Roma, Italy}

\textit{$^c$Perimeter Institute for Theoretical Physics, Waterloo,  Canada}
\end{center}

\vspace{1.5cm}

\begin{center}
\textbf{ABSTRACT}
\end{center}

{\leftskip=0.6in \rightskip=0.6in\noindent}
Over these past few years several quantum-gravity research groups have
been exploring the possibility that in some Planck-scale
nonclassical descriptions of spacetime one or another form of nonclassical spacetime
symmetries might arise. One of the most studied scenarios is based on the use
of Hopf algebras, but previous attempts were not successful
in deriving constructively the properties of the conserved charges one would like
to obtain from the Hopf structure, and this in turn did not allow
a crisp physical characterization of the new concept of spacetime symmetry.
Working within the example of $\kappa$-Minkowski noncommutative spacetime,
known to be particularly troublesome from this perspective, we observe that
these past failures
in the search of the charges originated from not recognizing the crucial role
that the noncommutative differential calculus plays in the
symmetry analysis.
We show that, if the properties of the $\kappa$-Minkowski differential calculus
are correctly taken into account, one can easily perform all
the steps of the Noether analysis and obtain an explicit formula relating
fields and energy-momentum charges. Our derivation also exposes the fact that
an apparent source of physical ambiguity in the description of the Hopf-algebra
rules of action, which was much emphasized in the literature,
actually only amounts to a choice of conventions and in particular does not
affect the formulas for the charges.

\newpage\baselineskip16pt plus .5pt minus .5pt \pagenumbering{arabic} %
\pagestyle{plain}

\addtolength{\topmargin}{-35mm}

\section{Introduction}
There has been quite some interest
recently (see, {\it e.g.}, Refs.~\cite{dsr1dsr2,jurekmax,leedsr})
in the hypothesis
that the short-distance (Planck-scale) structure of spacetime, which according to
a popular ``quantum-gravity intuition''~\cite{garay} should be highly nontrivial,
might be such to require a new description of spacetime symmetries.
In particular, the description of the ``Minkowski limit"~\cite{gacMinkLim}
of quantum gravity might
require some deformation of the Poincar\'e symmetries.
So far this idea has been mostly debated at a rather abstract conceptual level~\cite{gacMinkLim},
without the
support of a fully-worked-out theoretical picture that could at least illustrate
the type of phenomena to be expected from a deformation of
Poincar\'e symmetry.
One candidate nonclassical spacetime which several authors have considered
from this perspective is the $\kappa$-Minkowski noncommutative
spacetime~\cite{majrue,kpoinap},
with the characteristic space/time noncommutativity given by\footnote{The
space indices $j,l$
take values in $\{1,2,3\}$ while $0$ is the time index. We shall later also use the spacetime
indices $\mu,\nu,\alpha$, which take values in $\{0,1,2,3\}$.}
\bea
&[x_j,x_0]= i \lambda x_j \\
&[x_l,x_j]=0 ~,\label{kmnoncomm}
\eea
where the length\footnote{Rather than
our length scale $\lambda$ a majority of authors use the energy scale $\kappa$,
which is the inverse
of $\lambda$ ($\lambda \rightarrow 1/\kappa$).}
scale $\lambda$ is usually expected to be of the order of the
Planck length.
Some arguments based on ``mathematical analogies"\footnote{One notices
that $\kappa$-Minkowski
and the $\kappa$-Poincar\'e Hopf algebra form a ``Heisenberg
double"~\cite{lukiedouble,majidbook}, {\it i.e.} $\kappa$-Minkowski
and $\kappa$-Poincar\'e are linked, as algebras,
in a way that is rather similar to the relationship between classical Minkowski spacetime
and the classical Poincar\'e Lie algebra.}
would suggest that the symmetries of $\kappa$-Minkowski should be described in terms
of a $\kappa$-Poincar\'e Hopf algebra~\cite{majrue,kpoinap,lukieIW},
but it was never established whether
these formal observations are sufficient to ensure,
in the sense needed for physics applications,
the presence of nonclassical symmetries
governed somehow by the $\kappa$-Poincar\'e Hopf algebra.
In particular the conserved charges associated with
the $\kappa$-Poincar\'e (would-be-)symmetry transformations have never been obtained.
In the absence of an actual result the charges have been characterized on the
basis of various heuristic arguments, but these arguments lead to rather puzzling conclusions,
including the fact that the energy-momentum charges appear to be affected by an ambiguity:
a given field in $\kappa$-Minkowski would appear to carry different energy-momentum
charges depending on the choice of ordering convention made in describing the
field~\cite{kowaorder,aadluna}.
Other authors (see, {\it e.g.}, Ref.~\cite{kosiNOsymm})
have argued that the Hopf-algebra structures might after all not reflect the
presence of any symmetry:
the Hopf-algebra structures could be just a fancy mathematical formalization of
a rather trivial break down of symmetry.


We here attempt to bring the debate on Planck-scale-deformed spacetime symmetries,
at least in the $\kappa$-Minkowski framework, beyond heuristics. We show that previous
failures to derive energy-momentum conserved charges were due to the adoption of
a rather naive description of translation transformations, which in particular
did not take into account the properties of the noncommutative $\kappa$-Minkowski
differential calculus.
By taking properly into account the properties of the differential calculus
one encounters no obstruction in following  all
the steps of the Noether analysis and obtain an explicit formula relating
fields and energy-momentum charges.
We find four energy-momentum conserved charges using the invariance of the
theory under the four  $\kappa$-Poincar\'e translation transformations,
and this shows that Hopf algebras can be used to describe genuine spacetime symmetries.
The result we here derive confirms that in $\kappa$-Minkowski
there is a nonlinear Planck-scale modification of the
energy-momentum relation, but the nonlinearity intervenes in a way that differs
significantly from what had been conjectured on the basis of some heuristic arguments.
And it is noteworthy, in light of the mentioned debate on the possibility
of a puzzling ordering-convention dependence of the symmetry analysis
in $\kappa$-Minkowski, that our result expressing the energy-momentum charges as
functions of the field configuration does not depend in any way on the
choice of ordering convention for the description of the fields.

\section{Fields and translation generators}
Fields in $\kappa$-Minkowski (functions of the $\kappa$-Minkowski
noncommutative spacetime coordinates (\ref{kmnoncomm})) are
conveniently introduced~\cite{wessALL} in terms of a basis of
``Fourier exponentials": \be f(x)=\int d^4q \tilde{f}(\vec{q},q_0)
e^{i\vec{q} \cdot \vec{x}} e^{-i q_0 x_0} ~, \label{four} \ee
where the $q_\mu$ are ordinary commuting variables and  $\int
d^4q$ is an ordinary\footnote{For the $d^4q$ in (\ref{four}) one
may introduce\cite{gacmaj} an integration measure in order to
attribute certain desired transformation properties to the fields.
This will not play a role in our analysis.} integral. Therefore a
(noncommutative) field $f(x)$ is identified in terms of a
(commutative) field $\tilde{f}(\vec{q},q_0)$.

The type of description of fields given in (\ref{four}) suggests
straightforward generalizations to $\kappa$-Minkowski of some
familiar commutative-spacetime formulas. For example, one is
immediately led to a notion of integration\footnote{Besides the
integration over the $\kappa$-Minkowski coordinates one can also
describe the product rule for noncommutative fields (inherited
from the commutation relations among spacetime coordinates) in
terms of an equivalent deformed rule of product (a generalized
``Moyal star product"~\cite{wessALL}) for the associated
commutative fields. Within our analysis it is not too cumbersome
to work directly with the product of noncommutative fields, so we
skip the step of introducing the star product.} on the
$\kappa$-Minkowski coordinates by simply posing that \be \int d^4x
e^{i\vec{q} \cdot \vec{x}} e^{-i q_0 x_0} = \delta(\vec{q},q_0) ~,
\label{delta} \ee which leads to
\begin{eqnarray}
 \int f(x) d^4x = \tilde{f}(0,0) ~.
\label{integration}
\end{eqnarray}

It is much emphasized in the relevant literature~\cite{majrue,kowaorder,aadluna}
that translation generators, $P^{\mu}$,
in $\kappa$-Minkowski can be introduced with a ``classical action"
\be
P^{\mu}(e^{i\vec{q} \cdot \vec{x}} e^{-i q_0 x_0})
=q^{\mu}(e^{i\vec{q} \cdot \vec{x}} e^{-i q_0 x_0})  \label{PR}
\ee
(the action on any field is of course fully characterized, in light of (\ref{four}),
once the action on the exponentials $e^{i\vec{q} \cdot \vec{x}} e^{-i q_0 x_0}$ is given).

This type of classical-action description of translation generators
finds encouragement from many arguments~\cite{majrue,kowaorder,aadluna},
but it leads to a puzzle, which is exposed upon observing that one
may of course choose to describe our noncommutative fields
equivalently using different ordering conventions for the basis of
exponentials, such as
\be f(x)=\int d^4q
\tilde{f_{II}}(\vec{q},q_0) e^{-iq_{0}x_{0}/2}e^{i\vec{q}\cdot
\vec{x}}e^{-iq_{0}x_{0}/2}
 ~, \label{tsord}
\ee
which adopts a ``time-symmetrized ordering convention" instead
of the ``time-to-the-right ordering convention"
adopted in  (\ref{four}).
Since the $\kappa$-Minkowski commutation relations, (\ref{kmnoncomm}),
are such that $e^{i\vec{q}\cdot \vec{x}}e^{-iq_{0}x_{0}} =
e^{-iq_{0}x_{0}/2}e^{ie^{\lambda q_{0}/2 }\vec{q}\cdot \vec{x}}e^{-iq_{0}x_{0}/2}$
the same field $f(x)$ can indeed be equivalently described in terms of a
corresponding $\tilde{f}(\vec{q},q_0)$, according to (\ref{four}),
or in terms of $\tilde{f_{II}}(\vec{q},q_0)$,
according to (\ref{tsord}), and the two descriptions are simply related
by $\tilde{f_{II}}(\vec{q},q_0) =
e^{-3\lambda q_{0}/2 }\tilde{f}(e^{-\lambda q_{0}/2 } \vec{q},q_0)$.
But these two equivalent descriptions of fields lead to genuinely different ``classical
actions" of the translation generators. In fact, according to
the ``time-symmetrized ordering convention" one would introduce translation
generators through
\be
P_{II}^{\mu}(e^{-iq_{0}x_{0}/2}e^{i \vec{q}\cdot \vec{x}}e^{-iq_{0}x_{0}/2})
=q^{\mu}(e^{-iq_{0}x_{0}/2}e^{i \vec{q}\cdot \vec{x}}e^{-iq_{0}x_{0}/2})
 ~, \label{PRs}
\ee
and the $P_{II}^{\mu}$ are truly different from the $P^{\mu}$, as one sees by verifying
that
\be
P_{II}^{j}(e^{i\vec{q} \cdot \vec{x}} e^{-i q_0 x_0})
=e^{\lambda q_{0}/2 } q^{j}(e^{i\vec{q} \cdot \vec{x}} e^{-i q_0 x_0})
=e^{\lambda P_{0}/2 } P^{j}(e^{i\vec{q} \cdot \vec{x}} e^{-i q_0 x_0})
 ~. \label{PRverify}
\ee

This puzzling ``ordering ambiguity" is confined to the description of translation
generators. All other structures appear in fact to be independent of the choice of ordering
convention, including the description of rotation and boost generators (as shown in
Ref.~\cite{aadluna}) and the notion of integration
(whose independence on the choice of ordering follows from $\tilde{f}_{II}(0,0)=\tilde{f}(0,0)$).
But most of the interest in $\kappa$-Minkowski noncommutativity originates from
some expectations concerning its translation sector, and therefore this ordering ambiguity
has played an important role in the development of the relevant research area.
The most discussed candidate physical effect in $\kappa$-Minkowski
is a possible anomalous relation between energy and momentum (anomalous dispersion),
which could be interesting since it is conjectured to arise at a level that is not
far from the reach of forthcoming experimental
studies~\cite{grbgac,gampul,kifu,ita,gactp,jaco}.
But energy and momentum are the charges associated to translation invariance,
and the presence of alternative descriptions of the translation generators
is feared to introduce an ambiguity in the corresponding description of energy-momentum
charges.

\section{Translation transformations and differential calculus}
Having described the much-debated ``translations problem" for $\kappa$-Minkowski,
we now start introducing the first tools needed for our proposed solution of the problem.
Our first key observation takes as starting point another well-known characterization
of translation transformations, which rather than focusing on the generators concerns
the infinitesimal translation parameters. As customary in the commutative limit
one views an infinitesimal translation as a map $x_\mu \rightarrow x_\mu + \epsilon_\mu$.
When this concept is enforced in $\kappa$-Minkowski one
of course finds that the translation parameters must have nontrivial algebraic properties
\bea
[\epsilon_{j},x_0]= i \lambda \epsilon_{j}~,~~~
[\epsilon_{j}, x_{k}]=0~,~~~[\epsilon_{0}, x_{\mu}]=0 \label{kmdiffcalc}
\eea
in order to ensure that the ``point" $x+\epsilon$ still belongs to the $\kappa$-Minkowski
spacetime:
\bea
[x_j +\epsilon_{j},x_0+ \epsilon_0]= i \lambda (x_j+\epsilon_{j})~,~~
[x_i+\epsilon_{i},x_j +\epsilon_{j}]=0 ~.\label{check}
\eea
Of course these algebraic relations reflect the known
properties\footnote{As an alternative to the 4D differential calculus, whose properties
are reflected in the commutation relations (\ref{kmdiffcalc}), one may consider
a 5D differential calculus~\cite{Sitarz,Masl}. The artifact of a 5D differential calculus for
the 4D $\kappa$-Minkowski spacetime can be motivated~\cite{Sitarz,Masl}
on the basis of the desire to adapt the structure of the
differential calculus to some features of the rotation/boost sector of
the $\kappa$-Poincar\'e Hopf algebra. For our analysis, which concerns
translation symmetry, the 4D differential calculus should suffice.}
of the $\kappa$-Minkowski
differential calculus~\cite{majoeck} (the $\epsilon$'s describe the differences between
the coordinates of two spacetime points
and are therefore related to the $dx$'s of the differential calculus).

In order to perform the Noether analysis we must describe the action
of translation transformations on the fields $f$, which will be of the
type $f \rightarrow f+df$. It is crucial for our analysis to observe that the
two known facts about translations in $\kappa$-Minkowski, the form of the generators $P_\mu$
and the properties of the infinitesimal translation parameters $\epsilon_\mu$,
must be combined in the description of the $df$, as already
clearly encoded in the classical-spacetime formula $df= i [{P}^{\mu} f(x)] \epsilon_{\mu}$.
And the fact that in the $\kappa$-Minkowski case
the transformation parameters have nontrivial algebraic properties
confronts us with another ordering issue: as $df$ we could
take $i [{P}^{\mu} f(x)] \epsilon_{\mu}$
or, for example, $\{i [{P}^{\mu} f(x)] \epsilon_{\mu} +i \epsilon_{\mu} [{P}^{\mu} f(x)]\}/2 $.
There is clearly an infinity of different formulations of the $df$ which all reduce
to $df= i [{P}^{\mu} f(x)] \epsilon_{\mu}$ in the classical-spacetime (commutative) limit.

The definition of $df$ is however not to be treated as a freedom allowed by the formalism:
the exterior derivative operator $d$ must of course satisfy
the Leibnitz rule $d(f\cdot g)=f\cdot dg+df \cdot g$.\\
For the description of translations within the time-to-the-right ordering
convention, {\it i.e.} based on the translation generators $P^{\mu}$ of (\ref{PR}),
we considered the following {\it ansatz} for $df$
\be
df= i \left( \sum_{n} A_n \epsilon_{\mu}^{\alpha_{n}}
[P^{\mu}f] \epsilon_{\mu}^{1-\alpha_{n}} \right)  \label{ans}
\ee
where $A_n$ and $\alpha_n$ are real numbers, and $\sum_{n} A_n =1$
(meaning that we allowed $df$ to be written as a sum of terms with a variety of
possible ordering conventions for the position of the
transformation parameters with respect to the generators).
One then easily finds that the requirement (\ref{leib}) singles out the formula
\be
df= i \epsilon_{\mu}{P}^{\mu} f(x) ~. \label{trar}
\ee
It is through this formula, involving both generators and transformation parameters,
that one truly characterizes the translation transformations. The exclusive
knowledge of the properties of the translation generators is clearly insufficient.

Now that we have a genuine description of translation transformations, obtained working with
the time-to-the-right ordering convention, it is natural to ask whether a truly different
description of translation transformations is obtained adopting
the time-symmetrized ordering convention.
In order to investigate this issue we considered this alternative {\it ansatz}
for the $df$
\be
df_{II}= i \left( \sum_{n} B_n \epsilon_{\mu}^{\beta_{n}}
[P_{II}^{\mu} f] \epsilon_{\mu}^{1-\beta_{n}} \right)  \label{ansTWO}
\ee
formulated in terms of the generators $P_{II}^{\mu}$ encountered working
with the time-symmetrized ordering convention.
It is easy to verify that once again the requirement (\ref{leib}) leads to
a single possibility:
\be
df_{II}= i \epsilon_{\mu}^{1/2} [{P}_{II}^{\mu}  f(x)] \epsilon_{\mu}^{1/2} . \label{tras}
\ee
This turns out to be the way in which the formalism is telling us that there is a
unique concept of translation transformation (a unique $df$) in spite of the
availability of different choices of translation generators. In fact,
one easily verifies~\cite{antoPHD} that $df_{II}=df$.

\section{Noether analysis}
In the previous Section we obtained the needed starting point for the Noether
derivation of the conserved charges: the transformations for which we intend
to find associated conserved charges are now properly characterized in terms of
a map $f \rightarrow f+df$, rather than merely at the level of the generators. In doing this we accidentally
solved one of the most debated problems for $\kappa$-Minkowski theories: while at the
(insufficient) level of description based exclusively on the generators it appeared
that there would be an ambiguity in the definition of translation transformations,
we found that the algebraic properties of the $\kappa$-Minkowski translation
parameters are such that
different choices of formulation of the generators lead to
the same actual transformation (same formula for the associated $df$).
We now test our concept of translation invariance and our proposed generalization
of the Noether theorem within the most studied~\cite{kpoinap,kowaorder,aadluna}
 theory formulated
in  $\kappa$-Minkowski spacetime: a theory for a massless scalar field $\Phi(x)$
governed by the Klein-Gordon-like equation\footnote{This equation reduces to the
Klein-Gordon equation in the $\lambda \rightarrow 0$ limit, and its form
was proposed (see, {\it e.g.}, Refs.~\cite{kpoinap,kowaorder,aadluna})
using as guidance the idea that it should be an operator that commutes with all the
generators in the $\kappa$-Poincar\'e Hopf algebra.}
\be
C_{\lambda}(P_{\mu})\Phi \equiv  \left[ \left( \frac{2}{\lambda} \right)^2
 \sinh^2\left( \frac{\lambda P_0}{2} \right)-e^{\lambda P_0} \vec{P}^2 \right]\Phi=0
~,  \label{principale}
\ee
whose most general solution can be written as
\be
\Phi(x)=\int d^4k \tilde{f}(k_0, \vec{k})
e^{i \vec{k}\cdot \vec{x}}e^{-ik_0 x_0} \delta(C_{\lambda}(k_{\mu}))
 ~. \label{gensol}
\ee

The equation of motion (\ref{principale})
can be derived from the following action
\bea
S[\Phi]=\int d^4x \mathcal{L}[\Phi(x)]= \int d^4x {1 \over
2} \tilde{P}_{\mu}\Phi \tilde{P}^{\mu}\Phi
\label{Az}
\eea
where we introduced the compact notation $\tilde{P}_{\mu}$,
\bea
\tilde{P}_{0}=\left(\frac{2}{\lambda}\right)\sinh(\lambda {P}_{0}/2 ) \qquad  \tilde{P}_{j}
=e^{\lambda {P}_{0}/2} {P}_{j} ~,
\eea
which also allows to rewrite $C_{\lambda}(P_{\mu})$ as $\tilde{P}_{\mu}\tilde{P}^{\mu}$.\\
One easily finds that under a coordinate transformation $x \rightarrow x'$
the action varies according to
\bea
\delta S[\Phi] &\!\!\!=&\!\!\!  \int d^4x
\left( \mathcal{L}[\Phi'(x')] - \mathcal{L}[\Phi(x)] \right) =
- {1 \over
2} \int d^4x \left\{ e^{ \lambda P_0/2} \left[
( [\tilde{P}_{\mu} \tilde{P}^{\mu}) \Phi] \delta \Phi \right]
+ e^{- \lambda P_0/2} \left[ \delta \Phi ( \tilde{P}_{\mu} \tilde{P}^{\mu}) \Phi \right] \right\} + \nn \\
&&
 + \int d^4x \left\{ {1 \over
2} \tilde{P}^{\mu} \left[e^{ \lambda P_0/2} \tilde{P}_{\mu} \Phi \delta \Phi
+\delta \Phi e^{ -\lambda P_0/2} \tilde{P}_{\mu} \Phi \right]
+ \mathcal{L}[\Phi(x')] - \mathcal{L}[\Phi(x)] \right\}
 \label{doppiovar}
\eea In deriving (\ref{doppiovar}) it is useful to observe that
for the action of the operator $\tilde{P}_{\mu}$ on a product of
our noncommutative fields the following property holds: \be
\tilde{P}_{\mu} [f(x)g(x)]=[\tilde{P}_{\mu}f(x)]
[e^{\frac{\lambda}{2} P_0}g(x)] +[e^{-\frac{\lambda}{2} P_0}f(x)]
[\tilde{P}_{\mu}g(x)] \label{obs1} \ee
for any fields $f(x)$ and $g(x)$.\\
Clearly the terms in the first pair of curly brackets in (\ref{doppiovar})
simply reflect the fact that
this is indeed an action that generates (\ref{principale})
as equation of motion.
The terms in the second pair of curly brackets in (\ref{doppiovar})
should be used to obtain the form of the conserved currents
when $x \rightarrow x'$ is a symmetry transformation.



For our purposes it is necessary to analyze the variation of
action specifically under a translation transformation ($x
\rightarrow x+\epsilon$ and $\Phi \rightarrow \Phi+d\Phi$): \bea
\delta S[\Phi]&=& - \frac{1}{2} \int d^4 x \left[
(\tilde{P}_\alpha \Phi) ( \tilde{P}^\al i \epsilon^\mu
P_\mu\Phi)+(
\tilde{P}_\al   i \epsilon^\mu P_\mu \Phi)\tilde{P}^\al\Phi \right] +i \int d^4 x \epsilon^\mu P_\mu \mathcal{L}=  \nn\\
&=&- \frac{i}{2}\int d^4x \epsilon^\mu \left[ (e^{-\lambda P_0
\delta_{\mu j}}\tilde{P}_\alpha \Phi) ( \tilde{P}^\al P_\mu\Phi)+(
P_\mu\tilde{P}_\al\Phi)\tilde{P}^\al\Phi \right]+i \int d^4x  \epsilon^\mu P_\mu \mathcal{L}\nn\\
\eea where we used (\ref{trar}), the scalar-field transformation
properties of $\Phi$, and the observation that from
(\ref{kmdiffcalc}) one finds $\Phi \epsilon_j=\epsilon_j
e^{-\lambda \delta_{\mu j}P_0}\Phi$. Then using (\ref{obs1}) one
obtains \bea \delta S[\Phi]&=&-\frac{i}{2}\int d^4x \epsilon^\mu
\tilde{P}^\al [( e^{(-\delta_{\mu j}+\frac{1}{2})\lambda
P_0}\tilde{P}_\al \Phi) ( P_\mu\Phi)+( P_\mu \Phi) e^{-\lambda
P_0/2}\tilde{P}_\al\Phi ]+\nn\\
&&+\frac{i}{2} \int d^4x  \epsilon^\mu \Big\{( e^{(-\delta_{\mu
j}+\frac{1}{2})\lambda P_0} \tilde{P}_\al \tilde{P}^\al \Phi) (
 e^{\lambda P_0/2}P_\mu\Phi)+(  e^{-\lambda P_0/2}P_\mu\Phi)(
e^{-\lambda
P_0/2} \tilde{P}_\al \tilde{P}^\al \Phi)\Big\}+\nn\\
&&+i \int d^4x \epsilon^\mu P_\mu \mathcal{L} \label{cannavaro}
\eea
 We are of course interested in evaluating $\delta S[\Phi]$
for fields which are solutions of the equation of motion
(\ref{principale}), for which, since $\tilde{P}_\al \tilde{P}^\al
\Phi=0$, the term in curly bracket in (\ref{cannavaro}) vanishes.
It is easy to verify, also using the equation of motion and hence
\be \int d^4x e^{\xi P_0 } \big(f(x)g(x)\big)=\int d^4x f(x)g(x)
\ee that $\delta S$ can be reqritten in the form \bea \delta S=i
\int d^4x \left\{ \epsilon^\mu P^{\nu}J_{\nu\mu} \right\} ~,
\label{jocnew}\eea where \bea J_{j\mu}= {1 \over 2} (\tilde{P}_j
e^{(-\delta_{\mu j}+\frac{1}{2})\lambda P_0}\Phi) ( P_\mu\Phi)+ {1
\over 2} ( P_\mu \Phi)\tilde{P}_j e^{-\lambda
P_0/2}\Phi - \delta_{\mu j}P_j\tilde{P}_j^{-1}\mathcal{L} \nn\\
J_{0\mu}={1 \over 2} ( \tilde{P}_0 e^{(-\delta_{\mu
j}+\frac{1}{2})\lambda P_0}\Phi) ( P_\mu\Phi)+ {1 \over 2} ( P_\mu
\Phi)\tilde{P}_0 e^{-\lambda P_0/2}\Phi-\delta_{\mu
0}P_0\tilde{P}_0^{-1}\mathcal{L} \label{corre} \eea And by spatial
integration of the $J_{\mu\nu}$ one obtains as hoped four
time-independent quantities $Q_{\mu}$, the conserved charges. For
example for the $Q_j$ charges, \be Q_j=\int d^3x
J_{0j}=\frac{1}{2}\int d^3x [( \tilde{P}_0 e^{-\lambda P_0/2}\Phi)
( P_j\Phi)+( P_j \Phi)\tilde{P}_0 e^{-\lambda P_0/2}\Phi] ~, \ee
using a Fourier expansion of the field $\Phi$ solution of
(\ref{obs1}), one finds \bea Q_j&=&\frac{1}{2}\int d^4k
dp_0\phi(k)\phi(p_0,-ke^{\lambda k_0})e^{3\lambda k_0}\left[
\frac{1-e^{-\lambda p_0} }{\lambda}- \frac{e^{\lambda k_0}-1
}{\lambda}  \right] k_je^{i(k_0+p_0)x_0}
\delta\left(\left(\frac{2}{\lambda}\right)^2
\sinh^2\left(\frac{\lambda k_0}{2}\right)-e^{\lambda
k_0}k^2\right)\nn\\
&&\cdot \delta\left(\left(\frac{2}{\lambda}\right)^2
\sinh^2\left(\frac{\lambda p_0}{2}\right) -e^{\lambda (k_0+p_0)}
\left(\frac{2}{\lambda}\right)^2 \sinh^2\left(\frac{\lambda k_0}{
2}\right) \right) \label{qj} \eea whose time independence is most
easily seen by considering separately the two possibilities,
$(p'_0,k'_0)$ and $(p''_0,k''_0)$, allowed by the last delta
function in (\ref{qj}): \bea
p_{0}^{'}&=&-k_{0}^{'}\nn\\
e^{-\lambda p_{0}^{''}}&=&2-e^{\lambda k_{0}^{''}} \label{sol}
\eea Since the only possible sources of time dependence in
(\ref{qj}) are in factors of the form $e^{i(k_0+p_0)x_0}$ the
possibility $(p'_0,k'_0)$ does not give rise to any time
dependence. And for the possibility $(p''_0,k''_0)$ one easily
verified that the whole integrand vanishes.

The time independence of $Q_0$ can be verified
analogously~\cite{antoPHD}, and actually it is possible to
rewrite~\cite{antoPHD} all the charges in an explicitly
time-independent manner:
\be Q_{\mu}=\int d^3x J_{0\mu}=\int d^4p
\, {e^{3 \lambda p_0} \over 2} p_{\mu} \tilde{\Phi}(p_0, \vec{p})
\tilde{\Phi}(-p_0,- e^{\lambda p_0}\vec{p})
 \frac{p_0}{|p_0|}\delta(C_{\lambda}(p_{\mu}))
 ~. \label{cariche}
\ee

The fact that these energy-momentum charges $Q_{\mu}$
are indeed time independent confirms that
the Noether analysis has been successful.

And it is rather clear from the form of (\ref{cariche}) that the
energy-momentum relation is Planck-scale- ($\lambda$-)deformed
with respect to the special-relativistic (Poincar\'e-Lie-algebra)
limit. It is however also easy to see that for ``realistic" field
configurations, carrying energy much greater than the Planck
energy scale but obtained combining Fourier exponentials with
frequencies much lower than the Planck frequency, the Planck-scale
correction is always negligibly small. A simple way to
characterize this feature is found by considering a ``regularized
plane wave" solution~\cite{weinb}: \bea \Psi(x)=\frac{1}{2}
\frac{1}{\sqrt{|\vec{p}|V}}  \left( e^{i \vec{p}\cdot \vec{x}}
e^{-i \omega_{\lambda}x_0}+      e^{-i \vec{p} e^{\lambda
\omega_{\lambda}}\cdot \vec{x}} e^{i \omega_{\lambda} x_0} \right)
~,   \label{regularwave} \eea where $V$ represents a 3D
normalisation volume in the space-time and behaves as a regulator
and $\omega_{\lambda}(|\vec{p}|)$ stands for one of the two real
solutions of $C_{\lambda}(\omega_{\lambda},
\vec{p})=\left({2}/{\lambda}\right)^2 \sinh^2\left({\lambda
\omega_{\lambda}}/{2}\right)-\vec{p}^2e^{\lambda \omega_{\lambda}}
=0$. This would be a field with characteristic frequency scale
$\omega_{\lambda}$ that carries energy $Q_0 = \hbar
\omega_{\lambda}$. Substituting the field (\ref{regularwave}) in
the formulas (\ref{cariche}) one finds that \be
\left(\frac{2}{\lambda}\right)^2\sinh^2\left(\frac{\lambda
Q_0}{2}\right)-e^{\lambda Q_0}Q_i^2=0  \label{eureca} \ee Of
course, in the special-relativistic $\lambda \rightarrow 0$ limit
one recovers the standard energy-momentum relation $Q_0^2-Q_i^2=0$
(for our massless fields). For $\lambda \neq 0$ some corrections
are present, but these corrections quickly disappear if we
increase the intensity of the field. Indeed for the field
$\Psi_\alpha(x)=\alpha \Psi(x)$, obtained multiplying our
``regularized plane wave" $\Psi(x)$ by a real number $\alpha$, one
finds a $1/\alpha^2$ suppression of the correction, and in the
$\alpha \rightarrow \infty$ limit the dispersion relation regains
its special-relativistic form.

\section{Closing remarks}
The results here reported provide a safe point of anchorage
for the debate on the properties of energy-momentum charges
in $\kappa$-Minkowski.
Some of the properties conjectured on the basis of previous heuristic arguments
did emerge in our analysis, including the presence of some nonlinearity
in the energy-momentum relation for massless fields.
But the type of nonlinearity which emerged from our analysis
differs from all the forms that had been previously conjectured.
And we also showed
that some expectations based on those heuristic arguments are incorrect. In particular,
we found that the energy-momentum charges carried by a field do not
depend in any way on the choice of ordering convention
adopted in describing that field.

To our knowledge the characterization of translation symmetries in
$\kappa$-Minkowski that emerges from our Noether analysis is the
first explicit physical formulation of a non-classical
(``quantum") spacetime symmetry. The possibility that some sort of
non-classical spacetime symmetry could be relevant for
Planck-scale physics has been extensively discussed, but always
merely at the level of the properties of some algebras of would-be
symmetry generators, without establishing the properties of the
associated charges, and without ever really establishing whether
the new algebraic structures would result in something genuinely
new for some physical observables. Indeed, as mentioned in the
Introduction, some authors had argued that by introducing certain
types of new properties for the generators one might only be
providing fancy mathematics for structures which would not amount
to any new symmetry. Our translations in $\kappa$-Minkowski are
instead truly a new symmetry, and we have found that the fact that
the generators close on a Hopf algebra
leads to nontrivial properties for
some key physical observables (energy-momentum charges).

While it is significant, from a conceptual perspective, that such
a characterization of these Hopf-algebra spacetime symmetries is
finally available, our result does not necessarily provide support
for the idea that these symmetries should play a role in the
short-distance structure of spacetime. The presence of a
deformation of the energy-momentum dispersion relation is not in
itself too worrisome since there are independent
arguments~\cite{dsr1dsr2,jurekmax,leedsr} to motivate the
possibility of this feature in Planck-scale physics, but some
readers will understandably be puzzled by the emergence of a
``nonuniversal" dispersion-relation formula: as shown in the
previous section the energy-momentum charges of different fields
in $\kappa$-Minkowski are related by different dispersion
relations. And for realistically-large field configurations the
correction terms are negligibly small. So this first actual
(non-heuristic) encounter with a Hopf-algebra spacetime symmetry
is rather challenging at the conceptual level (by requiring that
we make sense of a nonuniversal dispersion relation, which the
theory in principle accommodates), and not much valuable from a
phenomenological perspective since for all practical purposes the
associated new effects are quantitatively irrelevant.

Future studies may explore whether
 something of greater value for phenomenology
is obtained if one manages to generalize our result
(which concerns
classical fields in our ``quantum" spacetime) to the case of quantum fields.
It seems plausible that, while classical fields are essentially unaffected by
the symmetry deformation, quantum particles in $\kappa$-Minkowski spacetime
be affected by a significant modification of the dispersion relation.
We have not yet attempted this generalization since at present one
finds in the literature several alternative
proposals~\cite{lukieFT,gacmich,jureFT}
of a quantum field theory in $\kappa$-Minkowski spacetime,
all unsatisfactory on one or another ground~\cite{gacMinkLim}.

Another possible direction for future studies is the one of considering
other noncommutative spacetimes.
It appears safe to assume that
our line of analysis is applicable to other noncommutative spacetimes, but
we were unable to propose a general recipe. The ingredients clearly should include
a Noether analysis and a proper combination of symmetry generators and transformation parameters;
however, the way in which we combined these ingredients made use
of some peculiarities of $\kappa$-Minkowski.
Probably the easiest generalization of our Noether analysis should apply to
other spacetimes which, like $\kappa$-Minkowski, are of ``Lie-algebra type"~\cite{wessALL}
($[x_\mu,x_\nu]=C_{\mu \nu}^\alpha x_\alpha$).
For spacetimes of ``canonical type" ($[x_\mu,x_\nu]=\theta_{\mu \nu}$)
the key issues are not in the translation sector but in the boost/rotation sector, and
this will perhaps require a bigger effort for the generalization of our procedure.

\end{document}